\documentclass[a4paper,aps,prb,preprintnumbers,amsmath,amssymb,twocolumn,superscriptaddress]{revtex4} %nofootinbib
 \pdfoutput=1 
\usepackage{amsmath}
\usepackage{amssymb}
\usepackage{verbatim}
\usepackage{graphicx}
\usepackage{color}
\usepackage{xspace}

 \newcommand{\ket}[1]{\left|#1\right\rangle} %|"cosa">
 \newcommand{\bra}[1]{\left\langle#1\right|} %<"cosa"|
\begin{document}
 
\title{Optical excitations in compressible and incompressible two-dimensional electron liquids}

\author{Tobias Gra{\ss}}
\affiliation{ICFO-Institut de Ciencies Fotoniques, The Barcelona Institute of Science and Technology, 08860 Castelldefels (Barcelona), Spain}
%\affiliation{Joint Quantum Institute, NIST and University of Maryland, College Park, Maryland, 20742, USA}
\author{Ovidiu Cotlet}
\affiliation{Institute of Quantum Electroncis, ETH Z\"urich, CH-8093 Z\"urich, Switzerland}
\author{Atac \.Imamo\u{g}lu}
\affiliation{Institute of Quantum Electroncis, ETH Z\"urich, CH-8093 Z\"urich, Switzerland}
\author{Mohammad Hafezi}
\affiliation{Joint Quantum Institute, NIST and University of Maryland, College Park, Maryland, 20742, USA}
\affiliation{Department of Electrical Engineering and Institute for Research in Electronics and Applied Physics, University of Maryland, College Park, MD 20742, USA}

\begin{abstract}
Optically generated electron-hole pairs can probe strongly correlated electronic matter, or, by forming exciton-polaritons within an optical cavity, give rise to photonic nonlinearities. The present paper theoretically studies the properties of electron-hole pairs in a two-dimensional electron liquid in the fractional quantum Hall regime. In particular, we quantify the effective interactions between optical excitations by numerically evaluating the system's energy spectrum under the assumption of full spin and Landau level polarization. Optically most active are those pair excitations which do not modify the correlations of the electron liquid, also known as multiplicative states. In the case of spatial separation of electrons and holes, these excitations interact repulsively with each other. However, when the electron liquid is compressible, other non-multiplicative configurations occur at lower energies. The interactions of such dark excitations strongly depend on the liquid, and can also become attractive. For the case of a single excitation, we also study the effect of Landau level mixing in the valence band which can dramatically change the effective mass of an exciton.
\end{abstract}
\maketitle

\section{Introduction}
Quantized electronic transport is the characteristic feature of integer and fractional quantum Hall systems \cite{klitzing80,tsui82}. It emerges when a two-dimensional electronic system is exposed to a strong perpendicular magnetic field. This intriguing transport behavior manifests the topological nature of the electrons' quantum state \cite{tknn82}, and the incompressibility of the topological liquid \cite{laughlin83}. To probe this physics beyond transport different methods of optical spectroscopy have been applied, including photoluminescence \cite{heiman88,turberfield90,goldberg90,buhmann90,yusa01,Byszewski06,nomura14}, inelastic light scattering \cite{pinczuk93,gallais08,levy16}, or absorption spectroscopy \cite{groshaus04,groshaus07,plochocka09}. Specifically, these techniques have enabled to study the spin physics of quantum Hall materials, including spin-wave excitations and topological spin textures (``skyrmions'') \cite{gallais08,levy16,groshaus04,plochocka09}. Recent advances have incorporated a quantum Hall system in an optical cavity \cite{smolka14,ravets18,knueppel19}. The formation of exciton-polaritons can lead to increased lifetimes of optical excitations, and optical nonlinearities have been detected using four wave mixing \cite{knueppel19}. Strikingly, interactions between exciton-polaritons were found to be strongly enhanced for some incompressible liquid phases, making such system a potential source for photonic non-linearities.

A simple theoretic model for optical excitations in a quantum Hall system restricts electrons and holes to the lowest Landau level (LLL) of the lowest subband, an assumption which holds for strong layer confinement and strong magnetic field. This model exhibits a remarkable ``hidden'' symmetry \cite{macdonald90,macdonald92,apalkov92}, making the optical excitations behave as an ideal Bose gas despite the presence of strong Coulomb interaction. In this scenario, the electron-hole pair has no explicit correlations with the electron liquid, and these excitonic states are therefore called ``multiplicative states''. They incur the system's entire oscillator strength, leading to a single emission/absorption line at a frequency which is independent on the filling factor. However, this remarkable theoretical result is not supported by experimental evidences from photoluminescence which show a non-trivial spectral structure, as for instance a doublet peak near filling $\nu=1/3$, cf.~Refs.~\onlinecite{goldberg90} or \onlinecite{nomura14}. This demonstrates that in real systems the hidden symmetry is broken, due to finite electron-hole separation in  asymmetric wells, and/or due to Landau level mixing. Theoretical attempts to explain the structure of emission spectra were made \cite{apalkov93,cooper97,Byszewski06,nomura14}, considering broken particle-hole symmetry in the interaction term.

However, the existing literature is limited mainly to the case of a single electron-hole pair. In the present paper, we go beyond earlier studies, as we also examine the behavior of a second pair. This approach allows to determine nonlinearities in optical experiments as a shift in the second pair's binding energy. At finite electron-hole separation, we obtain interacting states, amongst which some stand out due to a large overlap with the multiplicative states. We establish that these ``quasi-multiplicative'' states are the most relevant ones for optical experiments, although non-multiplicative configurations happen to be the ground state in compressible phases. Specifically, we show that energy differences between quasi-multiplicative states appear as the dominant peaks in photoluminescence spectra. Our numerical study of the system demonstrates that exciton-exciton interactions
are repulsive, but in contrast to the experiment of Ref.~\onlinecite{knueppel19} no dependence on the filling factor and/or the compressibility of the electron liquid is seen in the strength of the nonlinearity. This mismatch might be due to significant differences in the carrier density: By invoking the LLL approximation our theoretical study is valid for the high-density regime, becoming exact in the limit of infinite magnetic fields. In contrast, Ref.~\onlinecite{knueppel19} has been performed at rather low carrier densities, at which the lowest Landau level becomes fractionally filled in magnetic fields of only a few Tesla.

Our study is based on numerical diagonalization of the electron-hole Hamiltonian in a toroidal geometry \cite{asano02}. In contrast, the vast majority of the existing numerical work on electron-hole fluids, cf.~Refs.\onlinecite{macdonald90,macdonald92,apalkov92,apalkov93,apalkov95,wojs2000,wojs2000-2,Byszewski06,nomura14} has been performed on spherical surfaces. Like the sphere, the torus provides a compact geometry, but it is somewhat more realistic due to its equivalence to a rectangular plane with periodic boundary conditions. In particular, the rectangular model naturally allows for particle-hole symmetry breaking by confining electrons and holes to two parallel planes separated by a finite distance $d$. 

For the special case of particle-hole symmetry (i.e. within the lowest Landau level approximation and assuming spatially overlapping electron and hole layers), our study predicts a negative effective mass for the multiplicative exciton on top of a Laughlin liquid. In other words, the global ground state of the system occurs at finite momentum, and the momentum can be assigned to the electron-hole pair. Earlier numerical work on the sphere has seen a similar behavior, and has attributed it to the formation of a charged complex \cite{wojs2000,wojs2000-2}. By explicitly constructing trial wave function for the finite-momentum many-body states, we show that these states can rather be interpreted as dressed excitons, as in   Refs. \onlinecite{apalkov92,apalkov93,apalkov95}. We demonstrate that Landau level mixing as well as a finite distance between electrons and holes render the exciton mass positive. There, our account of Landau level mixing has been restricted to the valence band hole, as it is greatly enhanced due to the heavy mass of the hole.

The paper is organized in the following way: We describe our model of the system in Sec. II, and present the results in Sec. III. This section is sub-divided into three parts: The first part studies a system with a single pair excitation, the second part considers the system with two pairs. Both parts assume the lowest Landau level approximation, whereas in the third part we re-consider the scenario of a single pair excitation, but allowing for Landau level mixing in the valence band. A discussion which summarizes our results is given in Sec. IV. Technical details related to the numerical treatment of quantum Hall systems are given in the appendices.

\section{System and model}
We study electrons in a quantum well exposed to a strong perpendicular magnetic field $B$. To make the numerical treatment more tractable, we assume that both conduction and valence band electrons are spin-polarized, and the well confinement is strong enough to neglect subband mixing. The band structure is then given by flat Landau levels in  conduction and valence band. The energy gap between Landau levels is given by the cyclotron frequency  $\omega_B^\pm \equiv eB/m_{\rm eff}^\pm$, depending on the effective mass $m_{\rm eff}^\pm$ of the band, with index $+$ referring to the valence band, and index $-$ referring to the conduction band. Even for the extraordinarily light conduction band electrons in GaAs ($m_{\rm eff}^- \approx 0.07 m_{\rm e}$ with $m_{\rm e}$ the electron rest mass), the cyclotron gap  $\omega_B^-=2.5 {\rm THz} \times (B/{\rm T})$ is orders of magnitude smaller than the optical bandgap ($\hbar E_{\rm bg} \approx 2140 {\rm THz}$ in GaAs). Accounting for the valence band degrees of freedom in terms of holes, and switching into a frame which rotates with the bandgap energy, the single-particle Hamiltonian can be written as
\begin{align}
 H_0 = \hbar \sum_{n,j} \left( \omega_B^-e_{n,j}^\dagger e_{n,j} +   \omega_B^+ h_{n,j}^\dagger h_{n,j} \right),
\end{align}
Apart from a Landau level index $n$, the creation and annihilation operators for conduction band electrons ($e_{n,j}^\dagger, \ e_{n,j}$), and valence band holes ($h_{n,j}^\dagger, \ h_{n,j}$) carry a second index $j$.  Assuming the absence of disorder, this index is related to a gauge-dependent geometric symmetry of the system, e.g. rotational symmetry in the symmetric gauge or translational symmetry in the Landau gauge. For concreteness, we choose the latter one, in which the magnetic field is expressed through a vector potential ${\bf A}= B(0,-x)$, and thus $j$ is conveniently associated with invariant momentum along $y$, $p_y=\hbar k_y= \pm \hbar j \sqrt{\frac{2\pi \xi}{N_\Phi}}$, with opposite signs for electrons and holes. The spatial wave functions $\varphi_{n,j}(x,y)$ associated with these states are explicitly given in the appendix for a system with periodic boundary conditions (i.e. a torus), which have been chosen for this work.

In most parts of the present paper, we will apply the lowest Landau level (LLL) approximation, in which electrons and holes are restricted to level $n=0$. In this case, the single-particle Hamiltonian vanishes, and the interaction potential becomes the crucial Hamiltonian term. 
We consider a two-dimensional Coulomb potential for electrons and holes, but the planes to which different carrier types are confined may be different parallel layers spaced by a distance $d$. In Fourier space, the Coulomb potential then becomes $V(q)=\frac{1}{q}\exp(-d q)$, cf. Ref. \onlinecite{asano02,zhu16}. The divergent term at $q=0$ is excluded from the Fourier sum, which can be justified by assuming a homogeneous ``background'' charge density neutralizing each layer. However, in the real material charge neutrality applies only to the system as a whole, thus we need to add a charging energy $E_{\rm c}(N_{\rm h},d)$ which takes into account that each layer has a net charge $\pm N_{\rm h} e$. Accordingly, the charging term reads
\begin{align}
\label{echarge}
%E_{\rm c}(N_{\rm h},d)= 2\pi \frac{(N_{\rm h} e)^2 d}{\epsilon A}= \frac{N_{\rm h}^2 d/l_B}{N_\Phi} (e^2/\epsilon l_B).
H_{\rm c} = 2\pi \frac{e^2 d}{\epsilon A} N_{\rm h}^2 = \frac{d/l_B}{N_\Phi} (e^2/\epsilon l_B) N_{\rm h}^2,
\end{align}
where $A$ is the area of the system, $l_B=\sqrt{\hbar/eB}$ is the magnetic length, which is related to $A$ by the number of magnetic fluxes $N_\Phi$, $A=2\pi l_B^2 N_\Phi$.
As a convenient unit of energy, we use $e^2/\epsilon l_B$ throughout this paper, and $l_B$ as a unit for length.

The actual interactions are given through the Hamiltonian 
\begin{align}
\label{V}
 V &= \frac{1}{2} \sum_{\{n_i,j_i\}} \Big[ V_{j_1, j_2; j_3, j_4}^{n_1, n_2; n_3,n_4}(0) \big( e_{n_1,j_1}^\dagger e_{n_2,j_2}^\dagger e_{n_3,j_3} e_{n_4,j_4} + 
 \nonumber \\ & + h_{n_4,j_4}^\dagger h_{n_3,j_3}^\dagger h_{n_2,j_2} h_{n_1,j_1} \big) - 2 V_{j_1, j_2; j_3, j_4}^{n_1, n_2; n_3,n_4}(d) \times \nonumber \\ & \times e_{n_1,j_1}^\dagger h_{n_3,j_3}^\dagger h_{n_2,j_2} e_{n_4,j_4} \Big].
\end{align}
The interaction matrix elements $V_{j_1, j_2; j_3, j_4}^{n_1, n_2; n_3,n_4}(d)$ are evaluated in the appendix for the torus geometry. In the appendix, we also provide further details of our numerical study, in particular a discussion of the translational symmetry which leads to conserved many-body pseudomomenta. These provide quantum numbers for the many-body eigenstates \cite{haldane1985}, which we denote by integers $(K_x,K_y)$, defined modulo $N_\Phi$ and related to the pseudomomenta via $\tilde K_x \equiv K_x \frac{2\pi}{a}$ and $\tilde K_y \equiv K_y \frac{2\pi}{b}$. 

One advantage of periodic boundary conditions is the immediate and unique connection between particle-to-flux ratio and filling factor $\nu$, which characterizes the system in the thermodynamic limit. In the absence of holes, the filling factor is $\nu=N_{\rm e}/N_\Phi$. Charge-neutral optical excitations shall not change this value, and therefore we generalize the definition of the filling factor in the presence of $N_{\rm h}$ electron-hole pairs to:
\begin{align}
 \nu=\frac{N_{\rm e}-N_{\rm h}}{N_\Phi}.
\end{align}

At finite size, a system at a given Landau filling in the presence of a given number of electron-hole pair excitations is described by the numbers $N_{\rm e}, N_{\rm h}, N_\Phi$, and we denote its $i$th eigenstate at pseudomomenta ${\bf K}=(\tilde K_x, \tilde K_y)$ by $\ket{E_{N_{\rm e},N_{\rm h},N_\Phi}^{(i)}(K_x,K_y)}$, where $E_{N_{\rm e},N_{\rm h},N_\Phi}^{(i)}(K_x,K_y)$ stands for the energy of this state.
For convenience, we also define an exciton operator $X(k_x,k_y)^\dagger$ which connects a state of $N_{\rm e}$ electrons and $N_{\rm h}$ holes with a state of $N_{\rm e}+1$ electrons and $N_{\rm h}+1$ holes through addition of a pair at momentum $(k_x,k_y)$. The exciton operator is defined as
\begin{align}
\label{Xop}
 X(k_x,k_y)^\dagger = \sum_{j=0}^{N_\Phi-1} e^{i 2\pi j k_x/N_\Phi} e_{{\rm mod}(j+k_y,N_\Phi)}^\dagger h_j^\dagger.
\end{align}
If scattering into higher Landau levels is neglected, i.e. in the limit of an infinitely strong magnetic field, $X(k_x,k_y)^\dagger$ creates the exact vacuum excitations, i.e. the eigenstates at $N_{\rm h} = N_{\rm e}$. We use the ground state at $N_{\rm h} = N_{\rm e}=1$ to define the ``pure'' binding energy $E_X(N_\Phi,d)$ of an exciton:
\begin{align}
\label{EX}
E_X(N_\Phi,d) &= \frac{1}{N_\Phi} \sum_{j,j'} \langle {\rm vac} | h_{j'} e_{j'}  V e_j^\dagger h_j^\dagger | {\rm vac} \rangle + \frac{d}{N_\Phi}= 
\nonumber \\ &
= -\frac{1}{N_\Phi} \sum_{j,j'} V_{j', j; j', j}^{0,0;0,0}(d) + \frac{d}{N_\Phi}<0.
 \end{align}
In this expression, the term $\frac{d}{N_\Phi}$ accounts for the charging energy. In the thermodynamic limit,  $N_\Phi \rightarrow \infty$, the exciton binding energy converges to
\begin{align}
 E_X(d) &= -A \int \frac{d^2{\bf q}}{2\pi} V_{\bf q} \exp\left[-|{\bf q}|^2/2 \right] \nonumber \\ &= -\sqrt{\pi/2} \exp[d^2/2] {\rm erfc}[d/\sqrt{2}].
\end{align}

In contrast to the vacuum case, the exciton operator  $X(k_x,k_y)^\dagger$ does generally not produce exact eigenstates when acting on a state at fractional fillings, i.e. for $N_\Phi>N_{\rm e}>N_{\rm h}$. An exception occurs for overlapping conduction and valence bands, i.e. for $d=0$. Within the LLL approximation, the system then has 
 a ``hidden'' particle-hole symmetry \cite{macdonald90,macdonald92,apalkov92}, which formally is expressed by the commutator relation $[V,X(0,0)^\dagger]= E_X(N_\Phi,0) X(0,0)^\dagger$. This relation demands the existence of ``free'' excitonic states, created by $X(0,0)^\dagger$. The energies of these excitonic states are given by $E^{(i)}_{N_{\rm e}, N_{\rm h}, N_\Phi}(K_x,K_y) + E_X(N_\Phi,0)$. Repeated application of $X(0,0)^\dagger$ creates states with several free excitons, at energies $E^{(i)}_{N_{\rm e}, N_{\rm h}, N_\Phi}(K_x,K_y) + n E_X(N_\Phi,0)$, with integer $n$. As these states lack any correlations between the electron liquid and the additional electron-hole pairs, they are also called ``multiplicative states''. A major goal of our numerics in the following Section is to determine to which extent eigenstates at finite $d$ can be described in terms of such a multiplicative construction, and how much the true eigenstates are shifted from the energy levels of free excitons obtained via the multiplicative construction. Specifically, if these shifts scale non-linearly with the number of excitations in the system, this establishes an effective interaction between pairs.

\section{Results}
\subsection{Optical excitation within the LLL approximation}

\begin{figure*}[t]
 \centering
 \includegraphics[width=0.95 \textwidth]{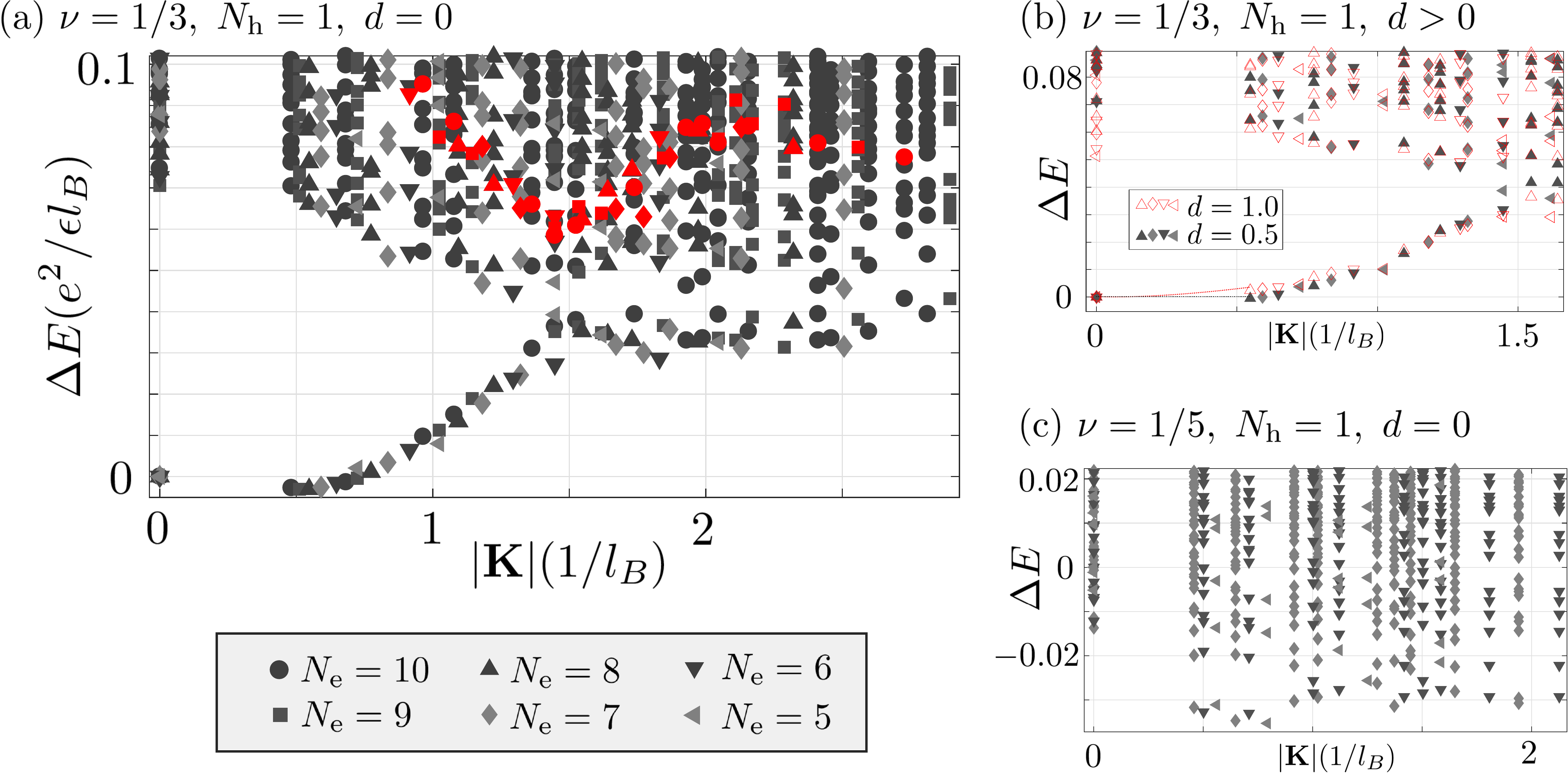}
 \caption{\label{fig:spec} {\bf Energy spectra with one electron-hole pair.} At Landau filling factors $\nu=1/3$ (a,b) and $\nu=1/5$ (c), we plot the energy spectra in the presence of one electron-hole pair for different system sizes (i.e. different electron numbers $N_{\rm e}$). In (a,c), we have chosen spatially overlapping conduction and valence bands ($d=0$), whereas (b) has separated bands. In all plots, the LLL approximation is assumed. We use the lowest multiplicative state at $K=0$ as an energy reference at each system size. In (a), the multiplicative magnetoroton states are plotted in red. In (b), the dashed lines between the lowest two states indicate the transition from infinite to finite positive effective exciton mass upon increasing the distance $d$.}
 \end{figure*}

Within the dipole approximation, the amplitude for optical interband transitions is proportional to the spatial overlap between the electronic wave functions in the two bands. This immediately leads to the selection rule $n,m \leftrightarrow n,m$, that is, conservation of Landau level and orbital quantum number \cite{schaefer-book}, and optical transitions are described by the operator $X^\dagger(0,0)$, introduced in the previous section. As mentioned there, an optical excitation obtained by acting with $X^\dagger(0,0)$ on an eigenstate of $V$ remains an eigenstates of $V$ in particle-hole symmetric systems (i.e. at $d=0$). In Fig. \ref{fig:spec}(a), we have identified these  ``multiplicative states'' within the full energy spectrum of an electron liquid at filling at $\nu=1/3$ in the presence of one electron-hole pair. In the absence of such a pair, the $\nu=1/3$ electron system is a strongly gapped incompressible liquid. Its ground state at $K=0$ is well described by Laughlin's wave function,  and the lowest (bulk) excitations are density modulations forming the magneto-roton branch, with a minimum at $|{\bf K}| l_B \approx \pi/2$. 

The multiplicative state originating from the Laughlin state, at $K=0$ and $\Delta E=0$ (i.e. this state  has been used as an energy offset in the plot), remains energetically separated from the bulk. In contrast, the multiplicative states originating from the magnetoroton branch, drawn in red in Fig. \ref{fig:spec}(a), are surrounded by many other energy levels. There is a well-defined excitation branch which connects the multiplicative Laughlin state with the bulk energy levels. Interestingly, this branch is found to be non-monotonic, with a global minimum at $K l_B \approx \pi/6$. In Ref. \onlinecite{wojs2000}, the states along this branch have been interpreted as charged complexes, but we note that the electron-hole correlation function does not show any accumulation of electrons in the vicinity of the hole, as compared to the neutral exciton state. Moreover, as seen from Table \ref{polovs}, these states can be modeled with reasonably good fidelity $F({\bf k}={\bf K})$ by acting with $X({\bf k})^\dagger$ from Eq. (\ref{Xop}) onto the Laughlin ground state.

\begin{table}
\begin{tabular}{|c|c|c|c|c|c|}
\hline
 $N_\Phi$ & $d_{\rm eh}$ & $F(0,0)$ & $F(1,0)=F(0,1)$ & $F(1,1)$ & $F(2,0)=F(0,2)$ \\
 \hline
 15 & 0   & 1      & 0.8537 & 0.7656 & 0.6503 \\
 15 & 0.5 & 0.9993 & 0.8776 & 0.8051 & 0.5526 \\
 \hline
 18 & 0   & 1      & 0.8679 & 0.7864 & 0.6368 \\
 18 & 0.5 & 0.9993 & 0.8883 & 0.8242 & 0.7021 \\
 \hline
 21 & 0   & 1      & 0.8784 & 0.8036 & 0.6785 \\
 21 & 0.5 & 0.9992 & 0.8982 & 0.8387 & 0.7328 \\
 \hline
\end{tabular}
\caption{\label{polovs}
For different momenta $(k_x,k_y)$, we list the fidelities $F(k_x,k_y)= |\langle E^{(1)}_{N+1,1,3N} | X^\dagger (k_x,k_y) | E^{(1)}_{N,0,3N} \rangle|$ of the multiplicative construction. The given numbers refer to filling factor $\nu=1/3$, at zero and at finite separation $d$ between electron and hole layers. Notably, the fidelity of the construction increases with system size.
}
\end{table}

These large fidelities indicate that the electronic correlations of the topological liquid are maintained by the optically excited system, supporting the notion of a dressed exciton branch. Also, from the electron-hole pair correlation function of these states we find that a single electronic charge is bound by the hole at both zero and finite momentum. However, as opposed to the case of a $K=0$ exciton, the charge distribution around the hole is not spherical-symmetric at finite momentum. In fact, these observations suggest to interpret the finite-momentum ground states as exciton-polarons \cite{sidler17,efimkin17,efimkin18}.

The non-monotonic behavior of this exciton-polaron branch renders the band's effective mass negative. This rather strange behavior is cured when electron and hole layers are at a finite distance $d$, as shown in Fig.~\ref{fig:spec}(b). At $d\approx 0.5 l_B$, the branch becomes monotonic. At this layer separation, the effective mass is infinite, as indicated by the horizontal black-dotted line in Fig. \ref{fig:spec}(b). For larger $d$, the effective mass becomes positive, cf. the red-dotted line in the plot. The (quasi-)multiplicative Laughlin state is then the true ground state of $\nu=1/3$ liquid in the presence of one electron-hole pair. Here, we have put the attribute ``quasi'' in parenthesis, because the $K=0$ state, while not being exactly the multiplicative state at finite $d$, stills has extremely large overlap with the multiplicative state ($>0.999$ at $d=0.5$, cf. Table \ref{polovs}). Anticipating a result from Sec. \ref{sec:mix}, we note that also Landau level mixing leads to positive effective exciton masses for any reasonable magnetic field strength.

In Fig.~\ref{fig:spec}(c), we show the spectrum of a system at $\nu=1/5$. As seen from Fig.~\ref{fig:mult}(a), at the given system size the pure electron system at $\nu=1/5$ lacks a gap, in stark contrast to $\nu=1/3$. In this context, we note that the $\nu=1/5$ Laughlin state, which is supported by a strong $V_3$ pseudopotential and which
in the thermodynamic limit of a Coulombic system is known to melt the surrounding crystallized phase \cite{goldmanWC90}, does not appear as a gapped ground state in finite-size studies \cite{chang05}. The spectrum in Fig.~\ref{fig:spec}(c) exhibits a large number of states at $\Delta E<0$ found at all momenta (including $K=0$).  As before, energies are measured from an offset defined by the energy of the multiplicative $K=0$ state. This finding demonstrates that in this gapless and/or compressible scenario the optically generated exciton will energetically be less favorable than for the incompressible liquid at $\nu=1/3$. We have checked that this holds true both for $d=0$ (shown in the plot), and at finite $d$ (not shown).

\subsection{Exciton-exciton interactions}

 \begin{figure*}[t]
 \centering
 \includegraphics[width=0.95 \textwidth]{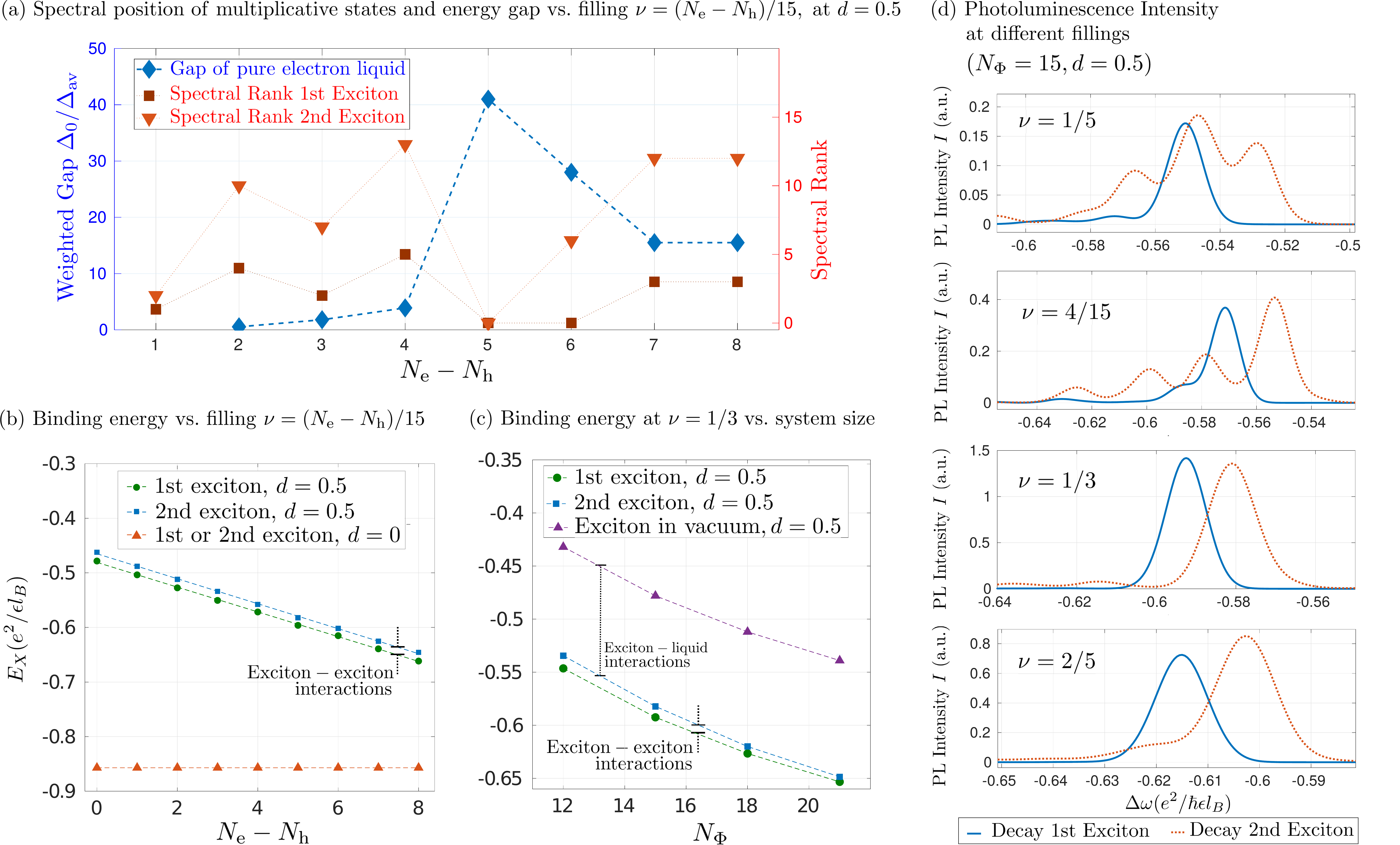}
 \caption{\label{fig:mult} {\bf Exciton-exciton interactions.} 
 (a) In blue: The energy gap $\Delta_0$ of a pure electron liquid ($N_{\rm h}=0$), weighted by the average level spacing $\Delta_{\rm av}$, is plotted as a function filling factor, i.e. as a function of $N_{\rm e}$ at fixed $N_\Phi=15$. Large gaps at $N_{\rm e}=5$ and $N_{\rm e}=6$ indicate incompressible behavior at fillings $\nu=1/3$ and $\nu=2/5$. In red: The spectral rank (at $K=0$) of the quasi-multiplicative states with one and two electron-hole pairs is plotted. Only at $\nu=1/3$ and $\nu=2/5$, the first multiplicative state (i.e. the multiplicative state with one pair) is ground state (spectral rank 0). The second multiplicative state for two pairs is the lowest-energy state only for $\nu=1/3$. (b,c) We plot the binding energy $E_X$ of the exciton in the first and the second multiplicative state as a function of filling factor (b), or system size (c).
 At $d>0$, different binding energies for the first and the second exciton indicate effective repulsive exciton-exciton interactions. These interactions turn out to be independent of the filling factor, and decreasing with system size. Compared to an exciton in the vacuum, see (c), $|E_X|$ is increased by an attractive interaction between exciton and the electronic liquid. (d) At different filling factors, we plot the frequency-resolved photoluminescence signal (measured as distance $\Delta \omega$ from the bandgap), assuming decay of the first or the second electron-hole pair. The distance between the peaks for the first and the second decay corresponds to the exciton-exciton interaction, and agrees with the values determined in panels (c,d) from the energies of the quasi-multiplicative states. This shows that the quasi-multiplicative states are the relevant states to determine the optical nonlinearities. The emission spectra are evaluated at an inverse temperature $\beta=100 \frac{e^2}{\epsilon l_B}$. This roughly corresponds to 2~K, if we choose a magnetic field $B=10~{\rm T}$ and dielectric constant $\epsilon_{\rm d}=12.9$ as for GaAs.
 }
 \end{figure*}

We further investigate the effect of compressibility (or ``gaplessness'') of the electron liquid on the behavior of multiple pair excitations. In recent four-wave mixing experiments \cite{knueppel19}, a quantum Hall system within an optical cavity has shown enhanced interactions between exciton-polaritons at certain filling factors which corresponded to  incompressible liquid phases (in particular at $\nu=2/5$). However, at other filling factors, including $\nu=1/3$ corresponding to the incompressible Laughlin liquid, no such effect has been seen. The mechanisms behind the enhancement remain unknown, and whether incompressibility generally leads to enhanced nonlinearities is an open question.
 
 In our numerical approach towards this question, we first collect a hint for incompressibility of the pure electron liquid by looking at the weighted energy gap $\Delta_0/\Delta_{\rm av}$ at different filling factors. The weighting is over the average level spacing $\Delta_{\rm av}$ at the given filling. The results are shown  in Fig.~\ref{fig:mult}(a): For the chosen system size ($N_\phi=15$), 
 incompressible behavior occurs at $\nu=1/3$ and $\nu=2/5$, in agreement with prominent fractional quantum Hall plateaux. Next, we analyzed the spectral rank of the first and the second quasi-multiplicative states, i.e. of those states which are obtained by acting once or twice with $X^\dagger(0,0)$ on the ground state of the liquid. The results are also presented in Fig.~\ref{fig:mult}(a): We find that only for the incompressible liquids (i.e. only at  $\nu=1/3$ and $\nu=2/5$) the ground state with one pair is given by the quasi-multiplicative state. Only for the Laughlin state ($\nu=1/3$), this is also true in the presence of a second pair. On the other hand, for all compressible liquids, the quasi-multiplicative state are always excited states. This generalizes our observation already made in the previous subsection in the context of the energy spectrum at $\nu=1/5$: Incompressibility of a liquid energetically favors the multiplicative construction as compared to other states. On the other hand, compressible liquids are able to find energetically more favorable ways to accommodate for electron-hole pairs than the formation of multiplicative excitonic complexes, e.g. through enhanced screening via polaron formation.
 
 Nevertheless, we continue our investigation by focusing on the energetic behavior of quasi-multiplicative states. This focus on quasi-multiplicative states is motivated by the important role of these states within optical setups. Specifically, we have evaluated the binding energy $E_X$ of the first and second multiplicative exciton. At $d=0$ and within the LLL approximation, as demanded by the hidden symmetry, the binding energy is independent from the number of electrons in the system, see Fig.~\ref{fig:mult}(b). However, at finite $d$, the binding energy is lowered due to the spatial separation between electron and hole. Since the exciton's finite dipole moment now allows for an effective interactions with the liquid, the binding energy becomes dependent on the filling factor (i.e. the density of the liquid). This interaction is found to be attractive, and thus leads to a monotonic increase of the binding energy with the density. We can quantify this exciton-liquid interaction by considering the difference to the binding energy of an exciton in the vacuum, as done in Fig.~\ref{fig:mult}(c). This plot also shows that the exciton-liquid interaction is independent from the system size. In both Fig.~\ref{fig:mult}(b) and Fig.~\ref{fig:mult}(c), we observe a mismatch of the binding energy for the first and the second exciton. This is a measure for an effective exciton-exciton interaction. This interaction is found to be repulsive, which naturally leads to a decay of interaction shifts with increasing system size, see Fig.~\ref{fig:mult}(c). On the other hand we note that the energy attributed to the exciton-exciton interaction is independent from the filling factor.
 
% This finding disagrees with the experimental results of Ref. \onlinecite{knueppel19}. Possible reasons for this discrepancy might be the ``idealizations'' of the theoretical model, which reduce the solid to two parabolic bands, without accounting for phonons and impurities. Moreover, our results are obtained within the LLL approximation and without taking into account the spin degree of freedom. However, 

One may speculate whether the quasi-multiplicative excitonic states are actually the right choice for determining the amount of exciton-exciton scattering. As discussed earlier, these states tend not to be ground state. If binding energies are calculated based on energy difference between the true ground states (with 0,1,2 electron-hole pairs), quite a different picture is obtained. Specifically, the binding energy difference between the first and second pair then depends of the filling factor, and it can even change its sign. While most filling fractions still yield repulsive exciton-exciton interactions, an energy shift corresponding to attractive interactions is found at $\nu=2/5$. This can be understood in the following way: From the spectral rank of the multiplicative  states, plotted in Fig. \ref{fig:spec}(a), we know that the ground state with one pair is a multiplicative state, whereas the second electron-hole pair is able to break the incompressibility of the liquid. This results in a lowering of energy, as compared to the energy of a second multiplicative exciton. If this lowering of energy is accounted for as an effective increase of the binding energy for the second pair, the second pair appears to be stronger bound than the first one. 

At this point, we are confronted with the question whether the excitonic states seen in optical experiments are described by the (quasi-)multiplicative states or by the state which accommodates for the electron-hole pair in the energetically most favorable way (i.e. the ground state). A figure of merit which answers this question in favor of the multiplicative state is the photoluminescence signal. Our quantitative photoluminescence model, which is further described in the appendix \ref{appPL}, assumes the decay from a thermal distribution over all states (including those at finite momentum), but with the number of optical excitations in the system being fixed. From this, we then obtain the frequency-resolved photoluminescence intensity shown in Fig.~\ref{fig:mult}(d) at different values for $\nu$. In our calculation, we independently consider two decay processes: One decay happens from a thermal state with one electron-hole pair, while the other process assumes a decay from a thermal state with two pairs. The relative shift of the two signals quantifies the interactions of bright excitons, and we obtain (for $d=0.5$) exactly the same result as we did before in Fig.~\ref{fig:mult}(c,d) by considering the energy of the multiplicative states.

As a side remark, we notice that, at $d=0.5$ and for incompressible fillings, almost no fine structure appears in the photoluminescence spectrum of a single decay channel. However, the combined measurement of different decay channels should exhibit some fine structure due to excitonic nonlinearities. Indeed, a splitting of the photoluminescence line has been observed in Ref. \onlinecite{Byszewski06}, and has been attributed to fractionally charged excitons. We note that the observed doublet splitting of about 0.4~meV is of the same order of magnitude as the excitonic nonlinearity within our theoretical model.

\subsection{Optical excitation with Landau level mixing \label{sec:mix}}

 \begin{figure}[t]
 \centering
 \includegraphics[width=0.47 \textwidth]{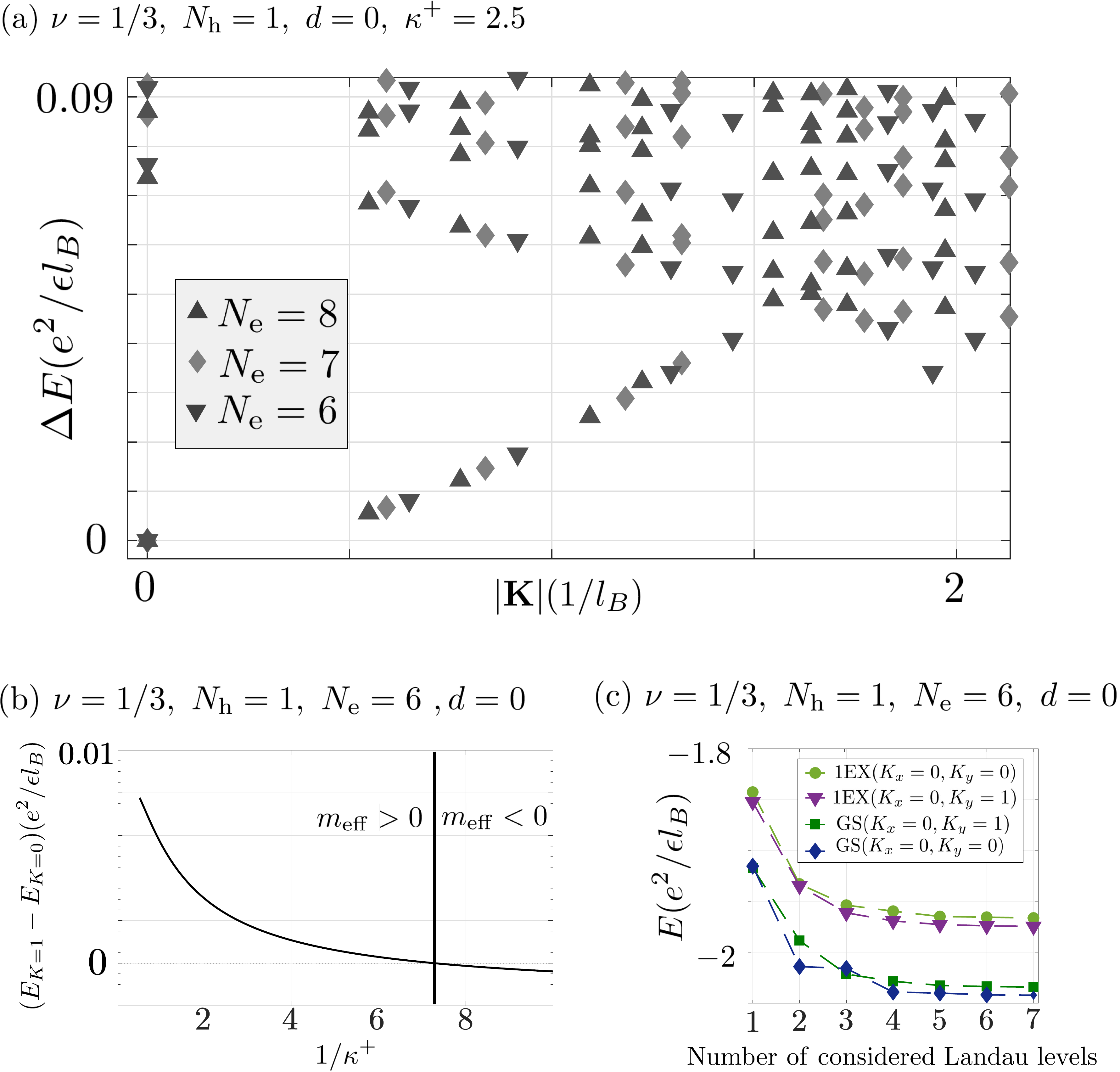}
 \caption{\label{fig:mix} {\bf Landau level mixing.} (a) Energy spectrum at $\nu=1/3$ in the presence of one electron-hole pair, taking into account Landau level mixing within the valence band. The chosen mixing parameter $\kappa^+=2.5$ corresponds roughly to $B=50{\rm T}$ in GaAs, a field strength at which Landau level mixing in the conduction band can safely be neglected. (b) Energy difference between the lowest state at $(K_x,K_y)=(1,0)$ and $(K_x,K_y)=(0,0)$ as a function of the mixing parameter $\kappa^+$. For $1/\kappa^+<7$ (or $B<13{\rm kT}$ in GaAs), the system enters in a phase with $E(K=1)-E(K=0)<0$, i.e. the effective mass $m_{\rm eff}$ becomes negative. (c) Convergence of the two lowest eigenvalues at $K=0$ and $K=1$ as a function of the number of valence band Landau levels which are taken into account. For the chosen mixing parameter, $\kappa^+=2.5$,  the error is kept below one percent when at least four Landau levels are taken into account. For the results in (a) and (b), we have considered six Landau levels.
 }
 \end{figure}

 The accuracy of the Landau level approximation made so far in this paper is controlled by the Landau level mixing parameter $\kappa^\pm$, the ratio of Coulomb energy versus Landau level spacing:
\begin{align}
\label{kappa}
 \kappa^\pm \equiv \frac{e^2}{\hbar \omega_B^\pm \epsilon l_B},
\end{align}
with $\pm$ distinguishing between valence and conduction band. Since $l_B\sim B^{-1/2}$, and  $\omega_B^\pm \sim B$, Landau level mixing tends to zero for large $B$,  $\kappa^\pm \sim B^{-1/2}$. However, even under an extremely strong magnetic field, e.g. $B=50~{\rm T}$, the lowest Landau level approximation turns out to be not well justified for holes in GaAs, $\kappa^+=2.4$, due to the holes' large effective mass, $m^+_{\rm eff} \approx 0.45~m_0$. In contrast, the light effective mass of conduction band electrons,  $m^-_{\rm eff}=0.067~m_0$, makes the lowest Landau level approximation quite a safe approximation for electrons, $\kappa^-=0.35$. As a function of the magnetic field, we get $\kappa^-=2.5/\sqrt{B[{\rm T}]}$ and $\kappa^+=16.7/\sqrt{B[{\rm T}]}$, where $B[{\rm T}]$ denotes the magnetic field strength in Tesla.

In the present section we will go beyond the LLL approximation. Quantitative improvements to a single Landau level approximation are possible by taking into account other Landau levels only virtually within a perturbative expansion \cite{bishara09,sodemann13}. However, this approach usually involves a decomposition of the Coulomb potential into pseudopotentials, which strongly affects the eigenvalues (in contrast to the rather weak effect of pseudopotential decomposition onto eigenstates). Alternatively, it is possible go beyond the single Landau level approximation by considering a Hilbert space which is increased by a finite amount of Landau level excitations \cite{wojs06}. The latter strategy is particularly well suited for our system of interest, as we may assume that Landau level mixing is restricted to the minority carriers (i.e. the holes).  Accordingly, we will consider the case of $N_{\rm e}$ electrons within the LLL, and a single hole, $N_{\rm h}=1$, for which a finite number $>1$ of Landau levels is admitted. Then, the Hilbert space dimension scales linearly with the number of Landau levels in the valence band, which allows us to take into account as many levels as needed for convergence. 

Qualitatively, the main effect of Landau level mixing is to destroy the hidden symmetry $[V,X(0,0)^\dagger]=E_X X(0,0)^\dagger$. It is not surprising that also the quantitative consequences of Landau level mixing are similar to the ones of a finite electron-hole separation, which breaks the hidden symmetry as well. Specifically, from the energy spectrum at $\nu=1/3$ plotted in Fig.~\ref{fig:mix}(a), we see that the ground state is shifted to $K=0$, in contrast to the finite-momentum ground state of the particle-hole symmetric system in Fig.~\ref{fig:spec}(a). As shown in Fig.~\ref{fig:mix}(b), the transition from the $K=0$ ground state into the finite-$K$ ground state occurs for 
$1/\kappa^+>7$ (i.e. for a gigantic field strength of $B>13{\rm kT}$ in GaAs). Thus, the scenario of an exciton negative effective masses is irrelevant from the experimental point of view.

The lowering of energy due to Landau level mixing in the valence band can be interpreted as an effective interaction between the exciton and the electron liquid. In fact, in the absence of a liquid, i.e. for an exciton on top of the vacuum, the ground state energy is not affect by Landau level mixing in the valence band. Even for $\kappa^+\rightarrow \infty$, no Landau level mixing occurs in the excitonic ground state, as long as $\kappa^-=0$ is kept at zero.

Fig. \ref{fig:mix}(c) allows to estimate the amount of Landau levels which need to be taken into account to accurately describe the system at the given mixing parameter $\kappa^+=2.5$. It is seen that the relative error in the eigenenergies is kept below 0.01 when at least four Landau levels are taken into account.
 
 \section{Summary}
 We have studied two-dimensional electron liquids in the quantum Hall regime in the presence of electron-hole pairs. Electron-hole pairs can  be generated optically, and can be used as a tool to probe the system, or to engineer photonic nonlinearities through the formation of exciton-polaritons. However, as our numerical work shows, multiplicative exciton states in which the electron-hole pair does not modify the correlations of the electronic liquid are not the energetically most favorable configurations at generic Landau filling factors, in particular those which correspond to compressible phases. Nevertheless, due to their large oscillator strength, these multiplicative exciton states are the most relevant states for optical experiments, and we have explicitly shown that the decay of these multiplicative states dominates the luminescence spectra. We also note that, if the system is embedded in an optical cavity, the large oscillator strength of these states causes a large AC Stark shift, which will make the exciton-polariton described by these states the ground state of the system. 
 
 From these perspectives, it seems justified to determine the strength of excitonic nonlinearities from the energy difference between quasi-multiplicative states with one and two excitons. In this way, we find a repulsive interaction between excitons, but the strength of these interactions shows no dependence on the filling factor. This result disagrees with recent experimental observations \cite{knueppel19} where some incompressible phases exhibit enhanced nonlinearities. This discrepancy may be due to the idealizations of our theoretical model in which we assume full spin polarization and disregard Landau level mixing. To be justified, these assumptions would require a very strong magnetic field.
 
 In this context, let us also emphasize the qualitative differences which we have obtained at zero layer separation ($d=0$) within and beyond the lowest Landau level approximation at filling $\nu=1/3$: Within the lowest Landau level approximation, the system exhibits a ground state at finite momentum, but Landau level mixing of the valence band hole leads to a zero momentum ground state. The finite-momentum ground state corresponds to an exciton with a negative effective mass. Finally, we note that excitonic nonlinearities might also be the cause for the broadening and/or splitting of luminescence line. In fact, the strength of the non-linearity between quasi-multiplicative states (obtained within the lowest Landau level approximation, but assuming a finite layer separation on the order of a few nm), is of the same order of magnitude as the splitting of the photoluminescence line seen experimentally \cite{Byszewski06}.

 \acknowledgments{The authors acknowledge discussions with Sina Zeytinoglu. 
TG acknowledges financial support from a fellowship granted by “la Caixa” Foundation (ID 100010434, fellowship code LCF/BQ/PI19/11690013), as well as funding
from the Spanish Ministry MINECO (National Plan 15 Grant: FISICATEAMO No. FIS2016-79508-P, SEVERO OCHOA No. SEV-2015-0522, FPI), European Social Fund, Fundaci\'o Cellex, Generalitat de Catalunya (AGAUR Grant No. 2017 SGR 1341 and CERCA/Program), ERC AdG NOQIA, EU FETPRO QUIC, and the National Science Centre, Poland-Symfonia Grant No. 2016/20/W/ST4/00314. MH acknowledges financial support from AFOSR FA9550-16-1-0323, and NSF Physics Frontier Center at the Joint Quantum Institute.}
 
 \appendix 

\section{ Exact diagonalization study}
In the following, we append some details regarding our description of the system and its numerical treatment. We have considered a truly two-dimensional electron-hole system in the Landau gauge with periodic boundary conditions. Here, we provide explicit expressions for the single-particle wave functions and the corresponding interaction matrix elements. Moreover, we discuss symmetries of the system.

\subsection{Single-particle wave functions}
In a gauge potential ${\bf A}\sim(0,x)$, the single-particle wave functions are plane waves along $y$-direction, and eigenstates of a harmonic oscillator along $x$.
The ground state level of the harmonic oscillator, given by a Gaussian $\exp(-\frac{1}{2}x^2)$, defines the lowest Landau level, while excited oscillator levels, obtained by multiplying the Gaussian with Hermite polynomials $H_n(x)$, yield the $n$th Landau level. The levels are equidistantly separated by a Landau level gap $\hbar \omega_B^\pm \equiv \hbar eB/m_{\rm eff}^\pm$, with $m_{\rm eff}^\pm$ being the effective masses of the positively and the negatively charged carriers. In each Landau level, there are $N_\Phi$ choices of a guiding center of the harmonic oscillator, $X_j=\frac{ja}{N_\Phi}$, with $j \in \{0,\dots,N_\Phi-1\}$. As the gauge potentials couples the $x$-coordinate to the momentum in $y$-direction, the guiding center also fixes the wavenumber of the plane wave. Periodicity in $x$-direction is obtained by summing over a periodic arrangement of guiding centers $X_j+ka$, with the summation in $k$ running from $-\infty$ to $\infty$. It is convenient to normalize length scales through the magnetic length, and account for the geometry of the system by a parameter $\xi \equiv a/b$. With this, $X_j=\sqrt{2\pi N_\Phi \xi}\frac{j}{N_\Phi}\equiv \alpha \frac{j}{N_\Phi}$, and the normalized wave functions can be written as \cite{yoshioka83}:
\begin{align}
\label{varphi}
 \varphi_{n,j}(x,y)= & \left(\frac{\xi}{2\pi^2N_\Phi}\right)^{1/4} \sum_{k=-\infty}^\infty \exp\left[i y  \alpha\left(\frac{j}{N_\Phi}+k\right) \right] \times
 \nonumber \\ &
 \exp\left[ -\frac{1}{2} \left(x- \alpha\left(\frac{j}{N_\Phi}+k\right) \right)^2 \right] \times
  \nonumber \\ &
 H_n\left[x-\alpha\left(\frac{j}{N_\Phi}+k\right)\right].
\end{align}
This wavefunction describes an electron in the $n$th Landau level. The quantum number $j$ quantifies its momentum in $y$-direction, $k_y=j \sqrt{\frac{2\pi \xi}{N_\Phi}}$ (in units $l_B^{-1}$).

\subsection{Interaction matrix elements}
Coulombic interactions occur between electrons and holes, but also with the nuclei, and, due to our choice of periodic boundaries, with mirror charges of each carrier. The latter can be neglected, since they only lead to a constant shift of all energy levels at a given torus ratio and given filling factor. The presence of nuclei make the system charge-neutral, and provide a homogeneous background potential given by $N_{\rm e}-N_{\rm h}$ positive charges. For convenience, we consider $N_{\rm e}$ positive background charges in the electronic layer, and $N_{\rm h}$ negative background charges in the hole layer, such that interactions with the background cancel the ${\bf q}=0$ contribution of the carriers' Coulomb potential, which would lead to divergent terms in the Fourier sums. As mentioned in the main text, this model does not take into account the fact that charge neutrality is only present in the system as a whole, and thus, for electrons and holes, we need to consider an additional charging energy. This energy contribution is given by Eq. (\ref{echarge}). 

In the following, we will evaluate the interaction matrix elements for the interactions between the carriers. The Fourier transform of the Coulomb potential reads \[V_d({\bf r}) = \pm \frac{2\pi}{A} \frac{e^2}{\epsilon} \sum_{\bf q} \frac{e^{i {\bf q}\cdot{\bf r}}}{|{\bf q}|} e^{-|{\bf q}|d},\]
where ${\bf q}$ is the in-plane wave vector. For $d=0$ and with the positive sign, the potential describes electron-electron interactions or hole-hole interactions, i.e repulsive interactions within a layer. For finite $d$ and with negative sign, the expression describes the electron-hole interactions, i.e. attractive interactions of opposite charge carriers confined to two layers separated by $d$.

The interaction matrix is given in Eq. (\ref{V}), with interaction matrix elements defined as
\begin{align}
  V_{j_1, j_2; j_3, j_4}^{n_1, n_2; n_3,n_4}(d) =& \frac{2\pi}{A} \frac{e^2}{\epsilon} \sum_{{\bf q} \neq 0} \frac{1}{|{\bf q}|}  \langle n_1,j_1 | e^{i {\bf q}\cdot {\bf r}} | n_4,j_4 \rangle
  \times \nonumber \\ &
  \langle n_2,j_2 | e^{-i {\bf q}\cdot {\bf r}} | n_3,j_3 \rangle e^{-|{\bf q}|d}.
  \end{align}
The position operator  ${\bf r}= {\bf R} + \delta {\bf r}$ can be decomposed into a guiding center ${\bf R}$ and a Landau orbit $\delta {\bf r}$, cf. Ref. \onlinecite{ezawa-book}. The guiding center is independent from the Landau level, and the corresponding matrix element can be evaluated in the lowest Landau level: $\langle n_1,j_1 | e^{i {\bf q}\cdot {\bf R}} | n_4,j_4 \rangle = \langle 0,j_1 | e^{i {\bf q}\cdot {\bf R}} | 0,j_4 \rangle$:
\begin{align}
 \langle 0,j_1 | e^{i {\bf q}\cdot {\bf R}} | 0,j_4 \rangle =&  \sum_{\Delta=-\infty}^{\infty} e^{-\frac{1}{4} (q_x^2+q_y^2)} e^{i\pi s (j_1+j_4+N_\Phi \Delta)} 
 \times \nonumber \\ &
 \delta_{t+j_1-j_4,N_\Phi\Delta}.
\end{align}
Here, $s$ and $t$ parametrize the quantized wavevector $(q_x,q_y)=\left(s\sqrt{\frac{2\pi}{N_\Phi\xi}},t,\sqrt{\frac{2\pi\xi}{N_\Phi}}\right)$.

The contribution from the Landau orbits is  $C_{n_1,n_4}(q_x,q_y) \equiv \langle n_1 | e^{i(q_x \delta x + q_y \delta y)} | n_4 \rangle$. To evaluate this, we note that the Landau orbits are related to the dynamical momentum $P_j =i\hbar \partial_j+e A_j$:
\[\delta x= - \frac{1}{eB} P_y \ \ \ {\rm and} \ \ \ \delta y = \frac{1}{eB} P_x.\]
These operators directly yield the Landau level raising and lowering operators:
\[ \hat a^\dagger = \frac{i l_B}{\sqrt{2}\hbar}(P_x+iP_y) \ \ \ {\rm and} \ \ \ \hat a = \frac{-i l_B}{\sqrt{2}\hbar}(P_x-iP_y). \]
Thus, with $q\equiv q_x-iq_y$, we can write $q_x \delta x + q_y \delta y = \frac{l_B}{\sqrt{2}}( q \hat a +  q^* \hat a^\dagger)$. Therefore,
$  C_{n_1,n_4}(q_x,q_y) = \langle n_1 | e^{i q \hat a / \sqrt{2}} e^{i q^* \hat a^\dagger / \sqrt{2}} | n_4 \rangle $. For $n_1 \geq n_4$, we get
\begin{align}
 C_{n_1,n_4}(q_x,q_y) =\sqrt{ \frac{ n_4!}{n_1!} } \left( \frac{ i q l_B}{\sqrt{2}} \right)^{n_1-n_4} L_{n_4}^{n_1-n_4}\left(\frac{(q_x^2+q_y^2)l_B^2}{2}\right)
\end{align}
For $n_1<n_4$, we use the relation $C_{n_1,n_4}(q_x,q_y) = C_{n_4,n_1}(-q_x,-q_y)^*$.

The interaction matrix elements are given by 
\begin{align}
 V_{j_1, j_2; j_3, j_4}^{n_1, n_2; n_3,n_4}(d) =& \frac{1}{N_\Phi} \frac{e^2}{\epsilon} \delta'_{j_1+j_2,j_3+j_4}\sum_{{\bf q} \neq 0} \frac{e^{-|{\bf q}|d}}{|{\bf q}|} 
 \times \nonumber \\ &
 C_{n_1,n_4}(q_x,q_y) C_{n_2,n_3}(-q_x,-q_y)  
  \times \nonumber \\ &
   \delta'_{j_1-j_4,t} e^{2\pi i s (j_1-j_3)} e^{-\frac{1}{2} (q_x^2+q_y^2)} 
\end{align}
The primed Kronecker symbols $\delta'$ are to be taken modulo $N_\Phi$.

Within the lowest Landau level approximation, all Landau level indices $n_i$ can be set to zero, and the interaction matrix elements reduce to
\begin{align}
  V_{j_1, j_2; j_3, j_4}^{0,0; 0,0}(d) =& \frac{1}{N_\Phi} \frac{e^2}{\epsilon} \delta'_{j_1+j_2,j_3+j_4}\sum_{{\bf q} \neq 0} \frac{e^{-|{\bf q}|d}}{|{\bf q}|}
 \times \nonumber \\ &
   \delta'_{j_1-j_4,t} e^{2\pi i s (j_1-j_3)}  e^{-\frac{1}{2} (q_x^2+q_y^2)}.
\end{align}

\subsection{Many-body basis and symmetries}
In finite-size studies, the full Hilbert space is characterized by $N_{\rm e},N_{\rm h},$ and $N_\Phi$. It becomes of finite dimension by assuming that only a finite number of Landau levels is relevant, and often, we even assume that the Landau level degrees of freedom are completely frozen (lowest Landau level approximation).
A many-body state is described by identifying the occupied single-particle states, i.e. by $(j_1^{\rm e},\dots,j_{N_{\rm e}}^{\rm e};j_1^{\rm h},\dots,j_{N_{\rm h}}^{\rm h})$ under the LLL assumption.

To diagonalize the Hamiltonian, we can greatly benefit from symmetries of the system. In the Landau gauge, the Hamiltonian is symmetric under (magnetic) translations. As seen already for the single-particle solutions, choosing the vector potential to be in the Landau gauge immediately leads to a conserved $y$-momentum. The Fock states are eigenstates of translation along $y$, and their $y$-momentum is obtained by summing the quantum numbers $j$ of occupied single-particle orbitals: 
\begin{align}
K_y = {\rm mod} \left( \sum_{i=1}^{N_{\rm e}} j_i^{\rm e} - \sum_{i=1}^{N_{\rm h}} j_i^{\rm h}, N_\Phi \right). 
\end{align}
The finite size of the system leads to equivalence between values $K_y$ differing by $N_\Phi$, so here we choose $K_y \in [0,N_\Phi-1]$. We note that $K_y$ is defined as an integer-valued quantum number, which corresponds to momentum $\tilde K_y \equiv K_y \frac{2\pi}{b}$.

To exploit the full translational symmetry \cite{haldane1985}, we need to construct a basis of eigenstates under magnetic translations also along the $x$-axis. 
For a filling factor $\nu=p/q$, with $p,q$ co-prime integers, we may consider the following set of Fock states: 
$\ket{f_0}\equiv(j_1^{\rm e},\dots,j_{N_{\rm e}}^{\rm e};j_1^{\rm h},\dots,j_{N_{\rm h}}^{\rm h})$,
$\ket{f_1}\equiv(j_1^{\rm e}+q,\dots,j_{N_{\rm e}}^{\rm e}+q;j_1^{\rm h}+q,\dots,j_{N_{\rm h}}^{\rm h}+q)$,
$\ket{f_2}\equiv(j_1^{\rm e}+2q,\dots,j_{N_{\rm e}}^{\rm e}+2q;j_1^{\rm h}+2q,\dots,j_{N_{\rm h}}^{\rm h}+2q) \dots$,
all of which are at the same momentum $K_y$. Invariant magnetic translations along $x$ are those which transform each member of this set into another member of the same set. Thus, eigenstates of these $x$-translations are constructed as a superposition of the $\ket{f_i}$, given by $\sum_r \exp(i [2\pi/(N_\Phi/q)] K_x r) \ket{f_r}$. The integer $K_x \in [0,N_\Phi/q-1]$ is recognized as a quantum number corresponding to pseudomomentum along $x$, $\tilde K_x = K_x \frac{2\pi}{a}$.

Using this construction, we divide the Hilbert space into blocks characterized by $(K_x,K_y)$. Additional symmetries leads to the equivalence between certain blocks:
Obviously, the system is invariant under center-of-mass (COM) translations. A COM translation along $x$ shifts the orbital of each carrier by some integer value $\Delta$: $j \rightarrow j+\Delta$. This transformation changes the momentum $K_y$ of a many-body state to $K_y+ \Delta (N_{\rm e}-N_{\rm h})\equiv K_y+\Delta N_\Phi/q$. Thus, each Fock state at $K_y$ is related to $q-1$ other Fock states at $K_y+\Delta N_\Phi/q$, with $\Delta=1,\dots,q-1$. Due to this equivalence between certain $K_y$-sectors, we  can restrict our study to a reduced Brillouin zone, where both $K_x$ and $K_y$ are restricted to $[0,N_\Phi/q-1]$. The Brillouin zone can further be reduced due to reflection symmetry and, for a square system, $C_4$ symmetry. Reflection symmetries lead to degenerate spectra at $K_x$ and $-K_x\equiv N_{\rm e}-N_{\rm h}-K_x = N_\Phi/q-K_x$, and $K_y$ and $-K_y\equiv N_\Phi/q-K_y$. The $C_4$-symmetry leads to degeneracies between $(K_x,K_y)$ and $(K_y,K_x)$.
For completeness, let us note that,  if $N_\Phi/q$ is even, there are two points $(K_x,K_y)=(0,0)$ and $(K_x,K_y)=(N_\Phi/2q,N_\Phi/2q)$ which are mapped onto themselves under reflection. We choose the origin of the Brillouin zone [i.e. the point $(K_x,K_y)=(0,0)$] in the sector of lower ground state energy, and, if needed, accordingly shift all pseudomomenta.

 \section{ Photoluminescence \label{appPL}}
The recombination of a $\Downarrow$ ($\Uparrow$) heavy hole and a $\uparrow$ ($\downarrow$) electron leads to emission of $\sigma^-$-polarized ($\sigma^+$-polarized) light. Within the dipole approximation, the envelope function of the electron/hole remains unchanged during a transition \cite{schaefer-book}, and the luminescence operator is given by $L=\sum_m e_m h_m$, cf. Refs. \onlinecite{macdonald92,apalkov93,wojs2000}. If an electron and a hole recombine in a system of $N_{\rm e}$ electrons and $N_{\rm h}$ holes, the resulting emission spectrum is given by
\begin{align}
\label{PLI}
 I_{N_{\rm e},N_{\rm h}}(\Delta\omega) =& \sum_{i,f}\delta(\hbar \Delta \omega+E_{N_{\rm e}-1,N_{\rm h}-1}^{(f)}-E_{N_{\rm e},N_{\rm h}}^{(i)}) \times
 \nonumber \\ & 
 P^{(i)}_{N_{\rm e},N_{\rm h}}(\beta) 
 \left|\bra{E^{(f)}_{N_{\rm e}-1,N_{\rm h}-1}}L \ket{E^{(i)}_{N_{\rm e},N_{\rm h}}}\right|^2 .
\end{align}
The argument of this function, $\Delta \omega$, is the difference of the photon frequency $\omega_{\rm ph}$ to the bandgap frequency $\omega_{\rm bg}$: $\omega_{\rm ph} = \omega_{\rm bg}+\Delta \omega$. The sum on the right-hand side of Eq. (\ref{PLI}) is over all states $i$ in the initial Hilbert space (i.e. before recombination), and all states $f$ in the final Hilbert space (i.e. after recombination). By $P^{(i)}_{N_{\rm e},N_{\rm h}}(\beta)$, we denote the thermal occupation of the initial states at an inverse temperature $\beta$: $P^{(i)}_{N_{\rm e},N_{\rm h}}(\beta) = \exp(-\beta E^{(i)}_{N_{\rm e},N_{\rm h}}) /{\cal Z}_{N_{\rm e},N_{\rm h}}(\beta)$ with ${\cal Z}_{N_{\rm e},N_{\rm h}}(\beta)=\sum_i \exp(-\beta E^{(i)}_{N_{\rm e},N_{\rm h}})$.
We note that, by assuming translational invariance or, equivalently, by neglecting disorder, transition matrix elements  $\left|\bra{E^{(f)}_{N_{\rm e}-1,N_{\rm h}-1}}L \ket{E^{(i)}_{N_{\rm e},N_{\rm h}}}\right|$ are zero, if initial and final state have different pseudomomenta. That is, by neglecting disorder we only account for direct interband transitions.

The photoluminescence spectrum is trivial if the model is particle-hole symmetric, that is for zero distance between electrons and holes, $d=0$, and within LLL approximation \cite{macdonald90,macdonald92}. In this limit, $[H,L] = E_X L$, and only the multiplicative states contribute to the emission spectrum with a resonance energy given by $E_X<0$,  independent from the electronic correlations.  The photoemission spectrum reduces to a single line. Non-trivial structure may only emerge when the hidden symmetry is broken (finite $d$ or Landau level mixing). As a technical remark, we note that we have artificially smoothened the spectral intensity in Fig.~\ref{fig:mult}(d) by replacing the Kronecker-$\delta$ in Eq. (\ref{PLI}) by a Gaussian of width $\sigma=5\times10^{-3} \frac{e^2}{\epsilon l_B}$. 
 
%\bibliography{bib}

\end{document}